# Observation of giant and tuneable thermal diffusivity of Dirac fluid at room temperature

Alexander Block[1,2], Alessandro Principi[3], Niels C.H. Hesp[1], Aron W. Cummings[2], Matz Liebel[1], Kenji Watanabe[4], Takashi Taniguchi[5], Stephan Roche[2,6], Frank H. L. Koppens[1,6], Niek F. van Hulst[1,6], Klaas-Jan Tielrooij[2,*]

[1]ICFO, The Institute of Photonic Sciences, The Barcelona Institute of Science and Technology, Castelldefels, Spain
[2]Catalan Institute of Nanoscience and Nanotechnology (ICN2), BIST & CSIC, Campus UAB, Bellaterra, Spain
[3]School of Physics and Astronomy, University of Manchester, Manchester, UK
[4]Research Center for Functional Materials, National Institute for Materials Science, 1-1 Namiki, Tsukuba 305-0044, Japan
[5]International Center for Materials Nanoarchitectonics, National Institute for Materials Science, 1-1 Namiki, Tsukuba 305-0044, Japan
[6]ICREA, Institució Catalana de Recerca i Estudis Avançats, Barcelona, Spain
*Correspondence to: klaas.tielrooij@icn2.cat

**Conducting materials typically exhibit either diffusive or ballistic charge transport. However, when electron-electron interactions dominate, a hydrodynamic regime with viscous charge flow emerges[1–13]. More stringent conditions eventually yield a quantum-critical Dirac-fluid regime, where electronic heat can flow more efficiently than charge[14–22]. Here we observe heat transport in graphene in the diffusive and hydrodynamic regimes, and report a controllable transition to the Dirac-fluid regime at room temperature, using carrier temperature and carrier density as control knobs. We introduce the technique of spatiotemporal thermoelectric microscopy with femtosecond temporal and nanometre spatial resolution, which allows for tracking electronic heat spreading. In the diffusive regime, we find a thermal diffusivity of ~2,000 cm$^2$/s, consistent with charge transport. Remarkably, during the hydrodynamic time window before momentum relaxation, we observe heat spreading corresponding to a giant diffusivity up to 70,000 cm$^2$/Vs, indicative of a Dirac fluid. These results are promising for applications such as nanoscale thermal management.**

In the diffusive transport regime, electrons and holes undergo random-walk motion with frequent momentum scattering events. When studied on ultrasmall length scales or ultrashort timescales, before momentum relaxation can occur, charges typically move ballistically – in a straight, uninterrupted line. Besides charge, electrons and holes carry electronic heat, with the thermal diffusivity linked to the charge mobility through the Wiedemann-Franz law. Under special conditions – when electron-electron interactions occur faster than momentum relaxation – the hydrodynamic regime emerges[1–22]. In this regime, the rapid collisions between particles can lead to viscous charge flow. The electron system then obeys macroscopic transport laws that are similar to the ones for classical fluid transport. During the last few years, signatures of viscous charge flow in this so-called Fermi-liquid regime were observed in 2D electron systems using electrical device measurements[3,4,10–12] and scanning probe microscopy[5,13,22] typically at cryogenic lattice temperatures and using ultraclean samples with high mobilities, μ > 100,000 cm$^2$/Vs.

A second hydrodynamic regime, which has no analogue in classical fluids, can occur very close to the Dirac point. When the Fermi temperature ($T_F = E_F/k_B$, where $E_F$ is the Fermi energy and $k_B$ is the Boltzmann constant) becomes small compared to the electron temperature $T_e$, the system becomes a quantum-critical fluid[6,9,14,15,17]. In this Dirac-fluid regime, the non-relativistic description of the viscous fluid is replaced by its ultra-relativistic counterpart, which accounts for the presence of both particles and holes, as well as for their linear energy dispersion. In line with theoretical predictions in this regime[15], noise thermometry measurements indicated a deviation from the Wiedemann-Franz law[19], electrical measurements of the thermoelectric Seebeck coefficient demonstrated a deviation from the Mott relation[20], and a terahertz-probe study revealed the quantum-critical carrier scattering rate[21]. Very recently, scanning probe measurements showed that viscous charge flow occurs both in the Fermi-liquid and Dirac-fluid regimes at room temperature[22]. However, observing and controlling the flow of electronic heat in the hydrodynamic regime at room temperature has so far remained elusive.

Here, we follow electronic heat flow in the diffusive and hydrodynamic regimes at room temperature, and demonstrate a controlled Fermi-liquid to Dirac-fluid crossover, with a strongly enhanced thermal diffusivity close to the Dirac point. These observations are enabled by ultrafast spatiotemporal thermoelectric microscopy, a technique inspired by all-optical spatiotemporal diffusivity measurements[23–25], with the crucial difference that the observable is the thermoelectric current, which is directly sensitive to electronic heat[26]. We use a hexagonal boron nitride (hBN)-encapsulated graphene device that is both a Hall bar for electrical measurements and a split-gate thermoelectric detector (Fig. 1a). Since we use ultrashort laser pulses, with an instrument response time of $\Delta t_{IRF}$ ~200 fs, to generate electronic heat, we are able to examine the system before momentum relaxation occurs, as we measure a momentum relaxation time $\tau_{mr}$ of ~350 fs (see Suppl. Fig. 1). In this temporal regime before momentum is relaxed, we enter the hydrodynamic window, because the electron-electron scattering time $\tau_{ee}$ is <100 fs [27], i.e. $\tau_{ee} < \Delta \tau_{IRF} < \tau_{mr}$. This is a different approach compared to most previous studies, where hydrodynamic effects were observed by using small system dimensions $L$ in order to eliminate effects of momentum scattering, i.e. $v_F \cdot \tau_{ee} < L < v_F \cdot \tau_{mr}$ [3–5,10–13,19,22] ($v_F = 10^6$ m/s is the Fermi velocity). Our approach furthermore exploits elevated carrier temperatures, which greatly increases the accessibility of the Dirac-fluid regime, as for increasing carrier temperatures the crossover occurs increasingly far away from the Dirac point[14,17] (see Fig. 1b). As we will show, during the hydrodynamic window significantly more efficient heat spreading occurs in the Dirac-fluid regime than in the Fermi-liquid regime and in the diffusive regime (see Fig. 1c-d).

Our technique works by employing two ultrafast laser pulses that produce localized spots of electronic heat within tens of femtoseconds[27]. These spots are characterized by an increased carrier temperature $T_e > T_l$, with $T_l$ the lattice temperature (300 K). The degree of spatial spreading of these electronic heat spots as a function of time is governed by the diffusivity $D$. We control the relative spatial and temporal displacement, $\Delta x$ and $\Delta t$, of the two pulses with sub-100 nm spatial precision and ~200 fs temporal resolution. Each laser pulse is incident on opposite sides of a *pn*-junction at a distance $\Delta x/2$ from the junction. This *pn*-junction is created by applying opposite voltages $\pm \Delta U$ with respect to the Dirac point voltages to the two backgates that form a split-gate structure. The two photo-generated electronic heat spots spread out spatially and part of the heat can reach the *pn*-junction after a certain amount of time, generating a thermoelectric current at the junction through the Seebeck gradient[26]. The small region of the *pn*-junction thus serves as a local probe of electron temperature. While each of the heat spots can create thermoelectric current independently, we obtain spatiotemporal information by examining exclusively the signal that corresponds to heat generated by one of the pulses interacting with heat generated by the other pulse – the interacting heat signal $\Delta I_{TE}$. Since the thermoelectric photocurrent scales sub-linearly with incident power, we can isolate this interacting heat signal $\Delta I_{TE}$ by modulating each laser beam at a different frequency, $f_1$ and $f_2$, and demodulating the thermoelectric current at the difference frequency $f_1 - f_2$. As illustrated in Fig. 1e-f, the higher the diffusivity *D*, the more interacting heat signal $\Delta I_{TE}$ remains for increasing $\Delta x$ and $\Delta t$.

Figure 2a shows the measured interacting heat current $\Delta I_{TE}$ as a function of $\Delta x$ and $\Delta t$. As expected, the largest $\Delta I_{TE}$ occurs for the largest spatiotemporal overlap at the *pn*-junction ($\Delta x = \Delta t = 0$). For increasing $|\Delta t|$, we find that the normalized signal extends further spatially, indicating the occurrence of heat spreading (see Fig. 2b). This spatial spread is quantified via the second moment $<\Delta x^2>$, which quantifies the width of the profile at different time delays (see Methods). Similar to recent all-optical spatiotemporal microscopy[24,25], we obtain spatial information beyond the diffraction limit by precise spatial sampling of diffraction-limited profiles. The experimentally obtained spatial spread as a function of $\Delta t$ (Fig. 2c) is very similar to the calculated results (Fig. 2d), obtained by simulating the experiment with a given diffusivity *D* (see Methods and Suppl. Note 1). The white lines indicate the values of the spatial spread $<\Delta x^2>$ for different $\Delta t$. We also compare the simulated spatial spread $<\Delta x^2>$ *vs.* $\Delta t$ (blue dashed line in Fig. 2e) with the theoretical expectation according to the heat diffusion equation, $<\Delta x^2> = <\Delta x^2>_{focus} + 2D\Delta t$ (dash-dotted line in Fig. 2e). Here, *D* is the same diffusivity that was used as input for the simulation, and $<\Delta x^2>_{focus}$ is the minimum second moment from the two overlapping pulses (see Suppl. Note 2 and Suppl. Figs. 2-5). The initial slope is the same for both the simulated heat spreading and the theoretical spreading following the heat diffusion equation.

We first discuss the experimental results in the diffusive regime, where $\Delta t > \tau_{mr}$. For three different gate voltage combinations, corresponding to Fermi energies between 75 and 190 meV ($T_F$ = 900 – 2200 K), we extract the spatial spread $<\Delta x^2>$ as a function of $\Delta t$ from $\Delta I_{TE}(\Delta x, \Delta t)$ (see symbols in Fig. 2e), and compare it with the results from simulations (solid lines). For these simulations, we have used the diffusivity values that we obtain directly from electrical measurements of charge mobility on the same device (see Suppl. Fig. 1), and the relation between mobility and diffusivity: $D = \mu E_F / 2e$ (see Methods). We find excellent agreement, if we account for short-lived ultrafast heat spreading around $\Delta t$ = 0, which leads to a larger-than-expected initial spread at time zero $<\Delta x^2>_{min}$, as we will explain below. Importantly, the agreement between the measured heat spread for $\Delta t > \tau_{mr}$ and the one calculated using the measured charge mobilities shows that electronic heat and charge flow together, as expected in the diffusive regime. Furthermore, it confirms that our technique is a reliable method for obtaining thermal diffusivities in a quantitative manner.

We now turn to the non-diffusive regime, by exploring the behaviour in the hydrodynamic window, where $\Delta t < \tau_{\mathrm{mr}}$. Surprisingly, the experimentally obtained spatial spreads start at a minimum value $<\Delta x^2>_{\mathrm{min}}$ larger than 2 µm², rather than starting at an expected $<\Delta x^2>_{\mathrm{focus}}$ = 0.56 µm². A second device we measured reproduces this larger-than-expected spatial spread at time zero (see Suppl. Note 3 and Suppl. Fig. 6). We exclude the possibility of an experimental artefact such as an underestimation of the laser spot size, since we repeated the measurements while scanning through the laser focus, and measured the focus size (Suppl. Figs. 2-5). Furthermore, we observe that the offset depends on the Fermi energy, while keeping all other experimental parameters fixed. We therefore attribute the large experimentally observed minimum $<\Delta x^2>_{\mathrm{min}}$ to ultrafast initial heat spreading that occurs before momentum relaxation takes place, $\Delta t \lesssim 350$ fs (see schematic illustration of spatiotemporal heat spreading in Fig. 1d). The dynamics of this initial heat spreading are washed out by the finite time resolution $\Delta t_{\mathrm{IRF}}$, and manifests as a large minimum $<\Delta x^2>_{\mathrm{min}}$ at time zero. The observed initial spatial spread suggests a thermal diffusivity of $D = (<\Delta x^2>_{\mathrm{min}} - <\Delta x^2>_{\mathrm{focus}})/2\Delta t_{\mathrm{IRF}} \approx 70{,}000$ cm²/s for the lowest measured $E_F$ of 75 meV. Simulations of heat spreading with an input diffusivity of 100,000 cm²/s are indeed consistent with the experimentally observed spread in the hydrodynamic window (see red line in Fig. 2e).

We attribute this observation of highly efficient initial heat spreading in the hydrodynamic time window to the presence of the quantum-critical electron-hole plasma. We can exclude that the observed initial spreading is the result of ballistic transport, as we calculate that the ballistic contribution to initial heat spreading would give only $<\Delta x^2>_{\mathrm{ball}}$ = 0.68 µm² (see Suppl. Note 2 and Suppl. Fig. 7). Besides, ballistic transport has a very weak dependence (<10 %) on carrier density in this range, as the Fermi velocity does not change significantly for the Fermi energies considered here[28].

In order to provide further evidence of hydrodynamic heat transport, we demonstrate the ability to control the crossover between the Fermi-liquid and quantum-critical Dirac-fluid regime via the ratio $T_e/T_F$, by independently varying $T_e$ via the incident laser power and $T_F$ via the applied gate voltages. A larger ratio results in less Coulomb screening and correspondingly stronger hydrodynamic effects due to electron-electron interactions. If $T_e$ is significantly larger than $T_F$, electrons and hole coexist, and the Dirac-fluid regime becomes accessible (see Fig. 1b). We perform spatial scans in the hydrodynamic window at a temporal delay of $\Delta t = 0$, in a geometry with one laser pulse impinging on the junction, while scanning the other pulse across (*x*-axis) and along (*y*-axis) the junction region. Figure 3a-d shows four representative spatial $\Delta I_{\mathrm{TE}}$ maps with varying $T_e/T_F$, yet similar signal magnitudes. Clearly, the signal is broader for larger $T_e/T_F$, indicating faster thermal transport. We repeat these measurements for a range of $T_e$ and $T_F$ values and quantify the initial heat spreading using Gaussian functions, with widths $\sigma_x$ and $\sigma_y$, to describe $\Delta I_{TE}$ at $\Delta t = 0$ as a function of $\Delta x$ or $\Delta y$ (see Fig. 3e-f and Suppl. Fig. 8). As expected for a crossover from the diffusive Fermi-liquid regime to the hydrodynamic Dirac-fluid regime, both spatial spreads $\sigma_x$ and $\sigma_y$ increase substantially for increasing ratio $T_e/T_F$. These spreads correspond to a diffusivity up to 40,000 cm²/s (see Methods), similar to the 70,000 cm²/s we found earlier.

We compare our experimental results to Boltzmann transport calculations following Refs. [9,18], including carrier interactions and long-range impurity scattering. We model impurities as Thomas-Fermi screened Coulomb scatterers of density $0.24 \cdot 10^{12}$/cm$^2$. Figure 3g shows the calculated thermal diffusivity $D$ as a function of $T_F$ and $T_e$, when considering only the hydrodynamic term due to electron-electron interactions, relevant in the hydrodynamic window where $\Delta t < \tau_{\mathrm{mr}}$. A higher electron temperature or lower Fermi temperature leads to strongly increased diffusivity, which is the same qualitative trend as for the experimental data taken at $\Delta t = 0$ in Fig. 3e-f, where a larger initial width originates from a larger diffusivity. The observed trend is clearly not consistent with calculations considering only the diffusive term due to scattering with impurities (Fig. 3h). The calculations thus support our interpretation of a crossover from a diffusive Fermi liquid to a hydrodynamic Dirac fluid for increasing ratio $T_e/T_F$.

A more quantitative comparison shows that the calculated $D$ in the diffusive regime is around 2000 cm$^2$/s (see Fig. 3h), in quantitative agreement with the experiment in the diffusive regime. The obtained thermal diffusivity in the hydrodynamic window close to the Dirac point reaches values above 100,000 cm$^2$/s, even higher than our experimental estimates of 35,000 – 70,000 cm$^2$/s. Using the calculated diffusivities, we estimate the spatial spread at time zero $\sigma_{\mathrm{calc}}$ (see Methods), as shown in Fig. 3g. These are similar to the experimentally obtained ones, thus confirming our conclusion of highly efficient heat spreading in the Dirac-fluid regime at room temperature, with a diffusivity that is almost two orders of magnitude larger than in the diffusive regime. We note that the theoretical calculations predict that even higher diffusivities are attainable.

Finally, we discuss the (3D) thermal conductivity, in order to assess the ability of the Dirac fluid to transport useful amounts of heat. We find ~100 W/mK in the diffusive regime (see Methods), in agreement with *ab-initio* calculations[29]. In the Dirac-fluid regime, with an electron temperature of ~1000 K, we obtain a thermal conductivity of 18,000 – 40,000 W/mK. This is in agreement with Ref. [15], where values up to 100,000 W/mK were predicted theoretically for large $T_e/T_F$. The thermal conductivity we obtain is about three orders of magnitude larger than the one obtained in the Dirac-fluid regime at cryogenic temperatures[19]. Interestingly, our results show that in the Dirac-fluid window the electronic contribution to heat transport can be much larger than the phononic contribution with a conductivity of >2000 W/mK [30], which is already exceptionally high. Thus, the Dirac electron-hole plasma can contribute very significantly to thermal transport, extracting heat from hot spots much faster than predicted by classical limits.

In conclusion, our results show that the, until recently unreachable, physical phenomena associated with the Dirac fluid do not only offer an exciting playground for interesting physical phenomena, yet also hold great promise for applications, *e.g.* in thermal management of nanoscale devices. We note that the quantum-critical behaviour can be switched on and off using a modest gate voltage and in systems prepared by standard fabrication techniques. Finally, we believe that the optoelectronic technique we have introduced will be a valuable tool to reach a better understanding of the thermal behaviour of a broad range of quantum materials, with great promise for novel technological applications.

**Methods**

*Fabrication of split-gate thermoelectric device*

The split-gate device with Hall geometry consists of exfoliated, single layer graphene encapsulated by hBN, using standard exfoliation and dry transfer techniques. The hBN-graphene-hBN stack is placed on a pre-defined split-gate structure made of graphene, grown by chemical vapour deposition, where the gap between the two gates is ~100 nm, created via electron-beam lithography and reactive ion etching (RIE). The top hBN and graphene are etched into a Hall bar shape with laser lithography and RIE, keeping the split-gate intact, and not etching completely through the bottom hBN. Finally, the Ti/Au side contacts are created by a further step of lithography, RIE and metal evaporation. The fabrication steps are shown in Supplementary Figure 9.

*Spatiotemporal thermoelectric current microscopy setup*

Our setup enables us to follow electronic heat spreading in space and time, because we use the thermoelectric signal generated by electronic heat interacting at a fixed location (the *pn*-junction), while we vary the spatial displacement of our two laser pulses with respect to this junction, and vary the temporal delay between the two ultrashort pulses. This means that we are following in space and time the diffusion of light-induced electronic heat from the location of light incidence to the *pn*-junction. It is the thermoelectric effect at the *pn*-junction, governed by the Seebeck coefficient, that generates our observable signal, the thermoelectric current. We note that although the value of the Seebeck coefficient itself changes when changing $E_F$, and when entering the hydrodynamic regime[20], this only affects the magnitude of the thermoelectric current – not how electronic heat is diffusing outside of the *pn*-junction, which is what we are following with our spatiotemporal technique.

A sketch of the setup is shown in Supplementary Figure 10. A Ti:sapphire oscillator (886 nm centre wavelength, 76 MHz repetition rate), is split into two beam paths. Both beams are modulated with optical choppers, at frequencies $f_1$ = 741 Hz and $f_2$ = 529 Hz. The relative time delay between the two pulses is controlled by a mechanical delay line. The spatial offset of one beam with respect to the other is controlled with a mirror galvanometer, while the position of the sample with respect to the beams is controlled with a piezo scanning stage. The beams are focused onto the sample with a 40x/NA 0.6 objective lens. We collect the TE photocurrent between the source and drain contacts on either side of the junction via lock-in amplification. This signal is measured by demodulation of the amplified current across the source and drain contacts through the graphene sheet. By demodulating the current signal at the difference frequency of the two modulation frequencies, $f_2 - f_1$ = 211.7 Hz, we isolate the signal caused by the interaction of both heating sources, which we call the interacting heat current $\Delta I_{TE}$. The temporal resolution of the setup of 200 fs is determined by the 20-80% rise time of transient absorption of graphene in the sample plane of the microscope (see Suppl. Fig. 11). The spatial accuracy is given by the signal-to-noise ratio, and is estimated to be below 100 nm.

We have used two distinct measurement geometries that each have their advantages and characteristics. For the data presented in Fig. 2, the two laser pulses are spatially offset symmetrically with respect to the gate junction region (by $\Delta x/2$ from the junction) by synchronized movement of the galvo mirrors (by $\Delta x$) and the piezo sample stage (by $\Delta x/2$). This measurement geometry is most suitable for extracting quantitatively the diffusivity, in particular in the diffusive regime, as shown by the simulations in Fig. 2. For the data presented in Fig. 3, where we focus on the hydrodynamic time window, we use a simpler "asymmetric" measurement geometry that gives a larger signal. Here, we keep one beam fixed on the junction while scanning the other beam by $\Delta x$, across the junction (Fig. 3e), and by $\Delta y$, along the junction (Fig. 3f), with fixed sample stage and moving the galvo mirrors only.

*Estimating Fermi temperature controlled by gate voltage*

During photocurrent measurements, the gate voltage $U_x$ is applied to the left (x = "A") or right (x = "B") side of the split-gate. We always apply a symmetric voltage around the experimentally determined Dirac point voltage $U_x^{DP}$: $U_A = U_A^{DP} + \Delta U$ and $U_B = U_B^{DP} - \Delta U$. The gate electrode and the graphene form a capacitor with the dielectric hexagonal boron nitride (hBN), with a thickness of $t_{hBN}$ = 70 nm, and a relative permittivity of $\epsilon_{hBN} = 3.56$. The carrier density $n$ is calculated via $n = \frac{\epsilon_0 \epsilon_{hBN}}{e\, t_{hBN}} \Delta U$, where $\epsilon_0$ is the vacuum permittivity. We calculate the Fermi energy $E_F$ and the Fermi temperature $T_F$ via $E_F^2 = \pi \hbar^2 v_F^2 \cdot n$, and $T_F = \frac{E_F}{k_B}$, where $k_B$ is the Boltzmann constant.

*Estimating carrier temperature controlled by laser power*

The thermoelectric photovoltage is assumed to be proportional to the time-averaged increase of the electronic temperature $T_e$ above the ambient temperature $T_0$, as in Ref. [31]. The sub-linear dependence of the thermoelectric current $I_{TE}$ on optical power for the device under study here for illumination with a single pulsed laser (λ = 886 nm) is shown in Supplementary Figure 12. With a linear temperature scaling of the electronic heat capacity for graphene away from the Dirac point, $C_e(T) = \gamma T$, we integrate the heat energy per unit area $dQ = C_e dT$, i.e., $\int_{Q_0}^{Q_0+\Delta Q} dQ = \int_{T_0}^{T_e} \gamma T dT$. With the incident power $P$ proportional to the absorbed heat energy per unit area $\Delta Q$, we find that the peak $T_e$ as a function of the laser power $P$ scales as[31] $T_e = \sqrt[2]{T_0^2 + bP}$. Here, the parameter b is defined via $bP = 2\Delta Q/\gamma$, and is used to convert incident power to peak electron temperature (see Suppl. Fig. 12).

*Simulation of the experiment*

A detailed description of the simulation can be found in Supplementary Note 1 and Supplementary Figure 13. In brief, we calculate the spatiotemporal evolution of electronic heat generated by the two optical pulses in the graphene sheet via the heat equation with a finite difference method. We define Gaussian heating pulses and calculate their temperature rise via the experimentally measured nonlinear power scaling. We extract the differential TE current contribution as a function of Δx and Δt by the difference of the heating at the *pn*-junction region in the presence of both pulses with respect to simulations with only one pulse at a time, analogous to the experimental difference-frequency demodulation.

*Quantifying the spatial spread*

The following analysis is performed both on the experimental data and on the simulated data of $\Delta I_{TE}(\Delta x, \Delta t)$ for "symmetric experiments" with optical pulses incident at a distance Δx on each side of the *pn*-junction (c.f. Fig. 1-2). For each Δt of the datasets $\Delta I_{TE}(\Delta x, \Delta t)$ we calculate the width of the signal via the second moment, which for an ideal Gaussian profile is equal to the squared Gaussian width $\sigma^2$. The second moment is calculated from the pixels $\Delta x_i$ (i = 1, ..., N) via

$$<\Delta x^2>(\Delta t) = \frac{\sum_i |\Delta x_i - \overline{\Delta x}|^2 \Delta I_{TE}(\Delta x_i, \Delta t)}{\sum_i \Delta I_{TE}(\Delta x_i, \Delta t)}, \text{ with the mean } \overline{\Delta x} = \frac{\sum_i \Delta x_i\, \Delta I_{TE}(\Delta x_i, \Delta t)}{\sum_i \Delta I_{TE}(x_i, t)}.$$

We note that the minimum second moment at the focus $<\Delta x^2>_{focus}$ of 0.56 µm² comes from simulating the symmetric experiment, using as input the measured Gaussian beam width at the focus $\sigma_{focus}^2$ = 0.14 µm² (see Suppl. Note 2). For the "asymmetric experiments" with one optical pulse always incident on the *pn*-junction (data of Fig. 3), we always consider the spatial profile only at time zero. Here we find that Gaussian fits with a background give the most reliable results. The entire set of data is shown in Supplementary Figure 8. For each dataset $\Delta I_{TE}(\Delta x)$ or $\Delta I_{TE}(\Delta y)$ taken at Δt = 0, we perform Gaussian fits using the function $f(\Delta x) = a \exp\left(-\frac{\Delta x^2}{2\sigma^2}\right) + b$, where the Gaussian squared width $\sigma^2$ indicates the thermal spreading. Here, the minimum simulated Gaussian widths are $(\sigma_x^2)_{focus}$ = 0.34 µm² and $(\sigma_y^2)_{focus}$ = 0.44 µm² (see Suppl. Note 2). The experimentally obtained widths from this dataset as function of gate voltage and optical power are also shown in Supplementary Figure 8, showing an increase with power, *i.e.* larger $T_e$, and an increase towards the Dirac point, *i.e.* smaller $T_F$. We estimate the theoretical Gaussian widths in Fig. 3g using $\sigma_{calc}^2 = (\sigma_x^2)_{focus} + 2D \Delta t_{IRF}$, where $D$ are the calculated diffusivities.

### Electrical measurements

We characterize our device electrically with four-probe measurements (see Suppl. Fig. 1), finding a charge mobility µ of 30,000 – 50,000 cm²/Vs, depending on carrier density. The measured mobilities correspond to a momentum scattering time $\tau_{mr}$ of 300 – 500 fs. Importantly, these scattering times are longer than the temporal resolution (instrument response function, IRF) of our measurement technique, $\Delta t_{IRF} \approx 200$ fs, thus allowing us to probe our system before and after momentum scattering occurs, *i.e.* in the non-diffusive and diffusive regime. We use these measured charge mobilities to calculate the expected thermal diffusivity via the Einstein relation[32,33] $\mu_{e/h} = \frac{e}{n_{e/h}} \frac{\partial n_{e/h}}{\partial E_F} D_{e/h}$, where $e$ is the elementary charge, $E_F$ is the Fermi energy, and $n_{e/h}$ is the electron/hole carrier density. For highly doped graphene ($E_F \gg k_B T$) the simple carrier density expression $n_{e/h} = \frac{E_F^2}{\pi \hbar^2 v_F^2}$, leads to the simple relation: $D_{e/h} = \frac{E_F}{2e} \mu_{e/h}$. We note that we obtain the identical result by calculating $D$ from the ratio of the 2D thermal conductivity $\kappa_{e,2D}$ and the electronic heat capacity $C_e$ and using the Wiedemann-Franz law: $\kappa_{e,2D}/\sigma = \pi^2/3 \cdot (k_B/e)^2 T_e$, where $k_B$ is the Boltzmann constant and $e$ the elementary charge, together with the conductivity $\sigma = ne\mu$ and the following heat capacity for graphene (valid for $T_e < T_F$): $C_e = \frac{2\pi \varepsilon_F k_B^2 T_e}{3\hbar^2 v_F^2}$. Given the measured mobilities, we expect thermal diffusivities around 2000 cm²/s for our sample.

### Thermal diffusivity and conductivity of the Dirac fluid

We estimate the enhanced thermal diffusivity of the Dirac fluid by comparing the measured width at time zero <$\Delta x^2$>$_{min}$ to the expected width <$\Delta x^2$>$_{focus}$ explained above, via $D$ = (<$\Delta x^2$>$_{min}$ - <$\Delta x^2$>$_{focus}$)/2$\Delta t_{IRF}$. We find values of 74,000 cm²/s for the symmetric scan (Fig. 2), and 29,000 cm²/s and 39,000 cm²/s for the x- and y-directions of the asymmetric scan (Fig. 3), where <$\Delta x^2$> is replaced with ($\sigma_x^2$) and ($\sigma_y^2$), respectively. The same calculation for a second device (see Suppl. Note 3 and Suppl. Fig. 6) gives a diffusivity of 100,000 cm²/s. The 3D thermal conductivity κ$_{3D}$ of the Dirac fluid is calculated from the diffusivity $D$ and the electronic heat capacity $C_e$, via κ$_{3D}$ = $DC_e/d$, where $d$ is the thickness of graphene, 0.3 nm. For the Dirac fluid, we have $T_e > T_F$, and therefore use the "undoped" electronic heat capacity[34] $\frac{18 \zeta(3)}{\pi(\hbar v_F)^2} k_B^3 T_e^2$, where $\zeta(3) \approx 1.202$. With the above estimate $D$ = 35,000 - 70,000 cm²/s and $T_e$ = 1000 K, we obtain the 3D thermal conductivity κ$_{3D}$ = 18,000 - 40,000 W/mK.

### Dirac fluid crossover temperature

Following the treatment in Ref. [14], we find the crossover temperature from Fermi liquid to Dirac fluid, as a function of Fermi temperature as $T_{cross}(T_F) = T_F \left(1 + \lambda \ln\left(\frac{T_0}{T_F}\right)\right)$, where $\lambda = e^2/16\epsilon_0 \epsilon_r v_F \hbar \approx 0.55/\epsilon_r$ for graphene with the dielectric environment $\epsilon_r \approx 3.56$ for hBN. The temperature $T_0 = \frac{2\hbar v_F \sqrt{\pi}}{3^{3/4} k_B a_0} \approx 8.4 \cdot 10^4$ K, with the inter-atomic distance $a_0 = 1.42 \cdot 10^{-10}$ m. The resulting crossover temperature is shown in Fig. 1b and as a red dashed line in Fig. 3e-g. We note that the relatively high refractive index of the hBN encapsulant makes the Dirac fluid more easily accessible, as it lowers the crossover temperature compared to vacuum, by a factor of about 2 for the range of $T_F$ studied here.

**Acknowledgements**

We thank Marco Polini and Pablo Piskunow for fruitful discussions. K.J.T., F.H.L.K., and S.R. acknowledge funding from the European Union's Horizon 2020 research and innovation programme under Grant Agreements No. 804349 (ERC StG CUHL), No. 785219 and No. 881603 (Graphene Flagship). K.J.T. also acknowledges funding from a RyC fellowship No. RYC-2017-22330 and IAE project PID2019-111673GB-I00. ICN2 was supported by the Severo Ochoa program from Spanish MINECO (Grant No. SEV-2017-0706). ICFO acknowledges financial support from Spanish MINECO through the "Severo Ochoa" program (SEV-2015-0522), from Fundació Privada Cellex, from Fundació Mir-Puig, and from Generalitat de Catalunya through the CERCA program. A.B. acknowledges financial support from the International PhD fellowship program 'la Caixa' - Severo Ochoa. A.P. is supported by the European Commission under the EU Horizon 2020 MSCA-RISE-2019 programme (project 873028 HYDROTRONICS). M.L. acknowledges support by the Ministry of Science, Innovation, and Universities (MCIU/AEI: BES-2016-078727, RTI2018-099957-J-I00 and PGC2018-096875-B-I00). K.W. and T.T. acknowledge support from the Elemental Strategy Initiative conducted by the MEXT, Japan, Grant Number JPMXP0112101001, JSPS KAKENHI Grant Number JP20H00354 and the CREST(JPMJCR15F3), JST. F.H.L.K. acknowledges financial support from the Government of Catalonia trough the SGR grant 1656. N.F.v.H. acknowledges financial support by the European Commission (ERC Advanced Grant 670949-LightNet), the Spanish MCIU (PGC2018-096875-B-I00) and the Catalan AGAUR (2017SGR1369).

**References**

1. Polini, M. & Geim, A. K. Viscous electron fluids. *Phys. Today* **73**, 28 (2020).

2. Müller, M., Schmalian, J. & Fritz, L. Graphene: A Nearly Perfect Fluid. *Phys. Rev. Lett.* **103**, 025301 (2009).

3. Gooth, J. *et al.* Thermal and electrical signatures of a hydrodynamic electron fluid in tungsten diphosphide. *Nat. Commun.* **9**, 4093 (2018).

4. Berdyugin, A. I. *et al.* Measuring hall viscosity of graphene's electron fluid. *Science* **364**, 162–165 (2019).

5. Sulpizio, J. A. *et al.* Visualizing Poiseuille flow of hydrodynamic electrons. *Nature* **576**, 75–79 (2019).

6. Foster, M. S. & Aleiner, I. L. Slow imbalance relaxation and thermoelectric transport in graphene. *Phys. Rev. B* **79**, 085415 (2009).

7. Narozhny, B. N., Gornyi, I. V., Titov, M., Schütt, M. & Mirlin, A. D. Hydrodynamics in graphene: Linear-response transport. *Phys. Rev. B* **91**, 035414 (2015).

8. Levitov, L. & Falkovich, G. Electron viscosity, current vortices and negative nonlocal resistance in graphene. *Nat. Phys.* **12**, 672–676 (2016).

9. Zarenia, M., Principi, A. & Vignale, G. Disorder-enabled hydrodynamics of charge and heat transport in monolayer graphene. *2D Mater.* **6**, 035024 (2019).

10. Moll, P. J. W., Kushwaha, P., Nandi, N., Schmidt, B. & Mackenzie, A. P. Evidence for hydrodynamic electron flow in PdCoO2. *Science* **351**, 1061–1064 (2016).

11. Bandurin, D. A. *et al.* Negative local resistance caused by viscous electron backflow in graphene. *Science* **351**, 1055–1058 (2016).

12. Krishna Kumar, R. *et al.* Superballistic flow of viscous electron fluid through graphene constrictions. *Nat. Phys.* **13**, 1182–1185 (2017).

13. Braem, B. A. *et al.* Scanning gate microscopy in a viscous electron fluid. *Phys. Rev. B* **98**, 241304 (2018).

14. Sheehy, D. E. & Schmalian, J. Quantum critical scaling in graphene. *Phys. Rev. Lett.* **99**, 226803 (2007).

15. Xie, H. Y. & Foster, M. S. Transport coefficients of graphene: Interplay of impurity scattering, Coulomb interaction, and optical phonons. *Phys. Rev. B* **93**, 195103 (2016).

16. Lucas, A., Crossno, J., Fong, K. C., Kim, P. & Sachdev, S. Transport in inhomogeneous quantum critical fluids and in the Dirac fluid in graphene. *Phys. Rev. B* **93**, 075426 (2016).

17. Lucas, A. & Fong, K. C. Hydrodynamics of electrons in graphene. *J. Phys. Condens. Matter* **30**, 053001 (2018).

18. Zarenia, M., Smith, T. B., Principi, A. & Vignale, G. Breakdown of the Wiedemann-Franz law in AB -stacked bilayer graphene. *Phys. Rev. B* **99**, 161407 (2019).

19. Crossno, J. *et al.* Observation of the Dirac fluid and the breakdown of the Wiedemann-Franz law in graphene. *Science* **351**, 1058–1061 (2016).

20. Ghahari, F. *et al.* Enhanced Thermoelectric Power in Graphene: Violation of the Mott Relation by Inelastic Scattering. *Phys. Rev. Lett.* **116**, 136802 (2016).

21. Gallagher, P. *et al.* Quantum-critical conductivity of the Dirac fluid in graphene. *Science* **364**, 158–162 (2019).

22. Ku, M. J. H. *et al.* Imaging viscous flow of the Dirac fluid in graphene. *Nature* **583**, 537–541 (2020).

23. Ruzicka, B. A. *et al.* Hot carrier diffusion in graphene. *Phys. Rev. B* **82**, 195414 (2010).

24. Zhu, T., Snaider, J. M., Yuan, L. & Huang, L. Ultrafast Dynamic Microscopy of Carrier and Exciton Transport. *Annu. Rev. Phys. Chem.* **70**, 219–244 (2019).

25. Block, A. *et al.* Tracking ultrafast hot-electron diffusion in space and time by ultrafast thermomodulation microscopy. *Sci. Adv.* **5**, eaav8965 (2019).

26. Gabor, N. M. *et al.* Hot carrier-assisted intrinsic photoresponse in graphene. *Science* **334**, 648–652 (2011).

27. Brida, D. *et al.* Ultrafast collinear scattering and carrier multiplication in graphene. *Nat.*

*Commun.* **4**, 1987 (2013).

28. Stauber, T. *et al.* Interacting Electrons in Graphene: Fermi Velocity Renormalization and Optical Response. *Phys. Rev. Lett.* **118**, 266801 (2017).

29. Kim, T. Y., Park, C. H. & Marzari, N. The Electronic Thermal Conductivity of Graphene. *Nano Lett.* **16**, 2439–2443 (2016).

30. Balandin, A. A. *et al.* Superior Thermal Conductivity of Single-Layer Graphene. *Nano Lett.* **8**, 902–907 (2008).

31. Tielrooij, K. J. *et al.* Out-of-plane heat transfer in van der Waals stacks through electron-hyperbolic phonon coupling. *Nat. Nanotechnol.* **13**, 41–46 (2018).

32. Zebrev, G. I. I. Graphene Field Effect Transistors: Diffusion-Drift Theory. in *Physics and Applications of Graphene - Theory* (IntechOpen, 2011).

33. Rengel, R. & Martín, M. J. Diffusion coefficient, correlation function, and power spectral density of velocity fluctuations in monolayer graphene. *J. Appl. Phys.* **114**, 143702 (2013).

34. Lui, C. H., Mak, K. F., Shan, J. & Heinz, T. F. Ultrafast photoluminescence from graphene. *Phys. Rev. Lett.* **105**, 127404 (2010).

**Figure 1.**

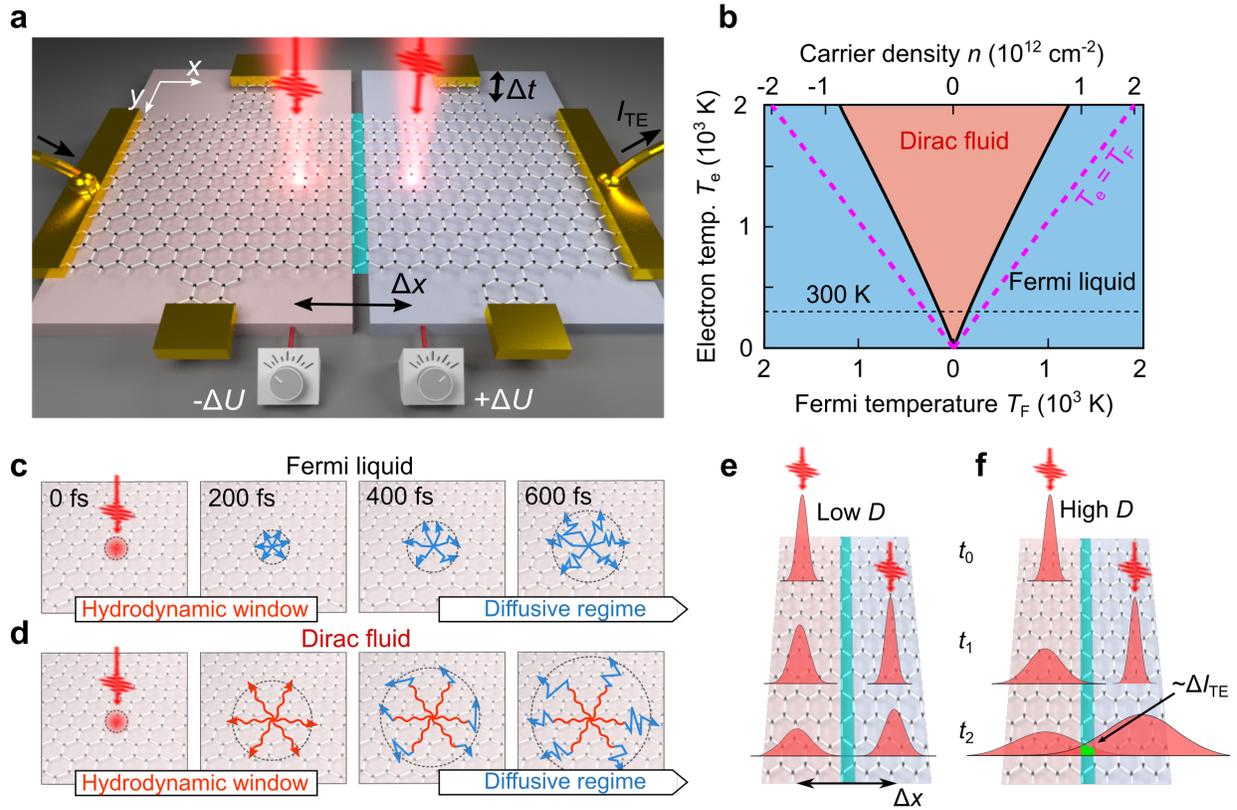

**Fig. 1. Spatiotemporal thermoelectric microscopy and heat spreading regimes.** (**a**) Concept of the experiment, where a graphene Hall-bar/thermoelectric device is illuminated by two femtosecond heat-generating pulses with a relative temporal offset Δ$t$ and a symmetric spatial offset Δ$x$ with respect to the *pn*-junction where electronic heat generates a thermoelectric current. The junction is created by applying +Δ$U$ to one gate and -Δ$U$ to the other. We isolate the differential thermoelectric current corresponding to light-induced electronic heat from both pulses that has travelled to the junction, where the heat adds up in a nonlinear fashion. (**b**) Phase diagram of the Dirac fluid regime, calculated following Ref. [14]. For increasing $T_e$ the Dirac-fluid regime occurs increasingly far away from the Dirac point. (**c, d**) Illustration of light-triggered spreading of electronic heat in the Fermi liquid regime (**c**) and Dirac fluid regime (**d**). In both cases, for Δ$t > \tau_{\mathrm{mr}}$, diffusive transport dominates (straight blue lines), while in the hydrodynamic window, with Δ$t < \tau_{\mathrm{mr}}$, extremely efficient heat transport occurs in the Dirac-fluid regime (wavy red lines). (**e, f**) Sketch of the spatial broadening of the heat spots for low (**e**) and high (**f**) diffusivity, indicating a higher interacting heat at the junction region, hence higher Δ$I_{\mathrm{TE}}$ signal for higher $D$.

**Figure 2.**

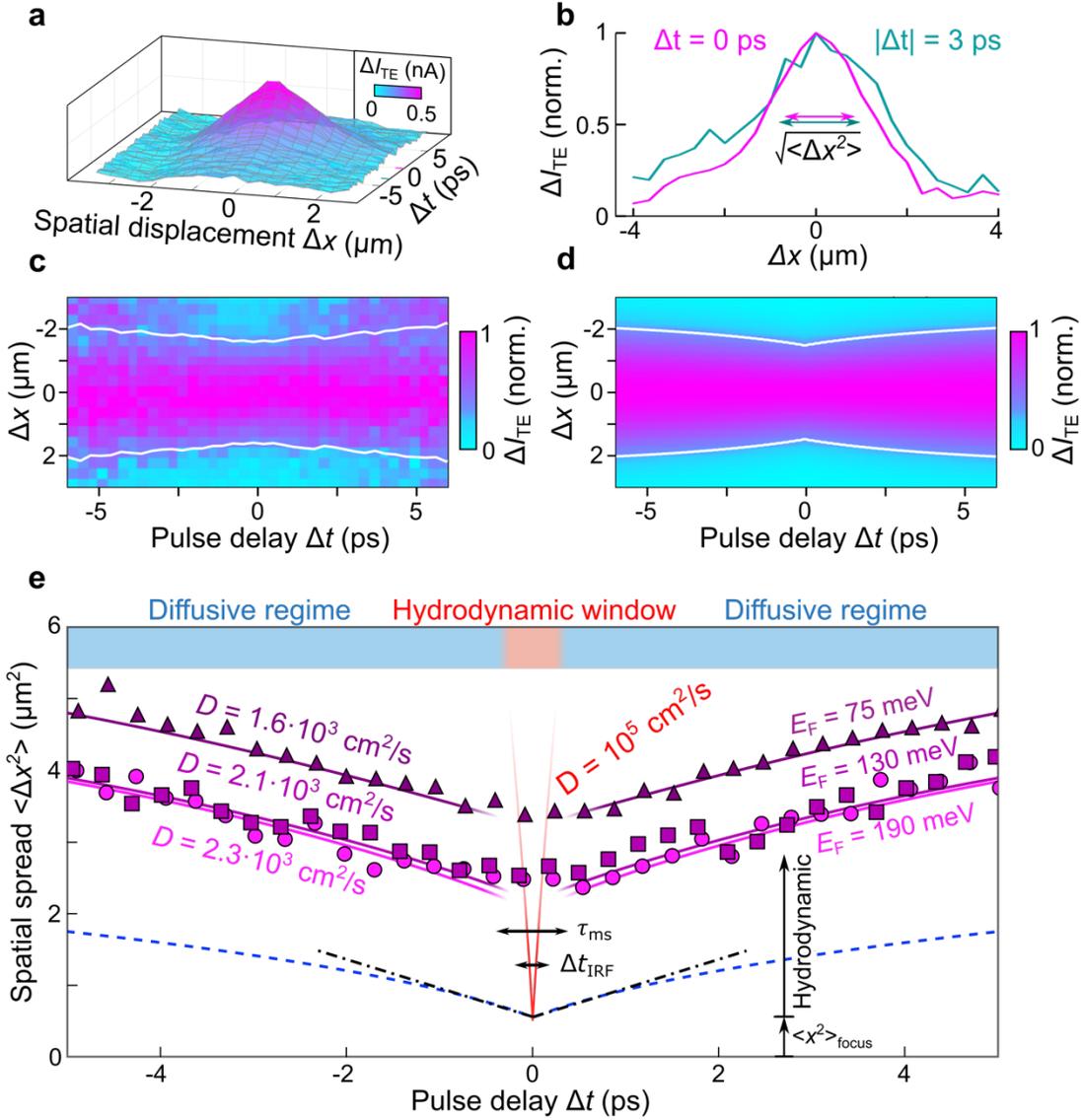

**Fig. 2. Spatiotemporal tracking of heat spreading.** (**a**) The experimental spatiotemporal differential thermoelectric current $\Delta I_{TE}$ as a function of $\Delta x$ and $\Delta t$. (**b**) Normalized profiles, showing a larger spatial extent for larger $|\Delta t|$. (**c, d**) Experimental (**c**) and simulated (**d**) normalized $\Delta I_{TE}$ for each $\Delta t$, showing spatial broadening due to thermal transport as a function of $\Delta t$. The white line indicates the spatial spread $<\Delta x^2>$. (**e**) Spatial spread $<\Delta x^2>$ of $\Delta I_{TE}$, as a function of $\Delta t$ for three different Fermi energies (symbols), with simulation results using as input the diffusivities from electrical mobility measurements (purple solid lines), with offset due to ultrafast heat spreading around time zero. Simulation (blue dashed line) and theoretical heat equation (black dash-dotted line) results with the same input diffusivity and no ultrafast spreading around time zero. Heat spreading with ultrahigh diffusivity in the Dirac-fluid regime (red line), which lasts for a few hundred fs, explains the time zero offset.

**Figure 3.**

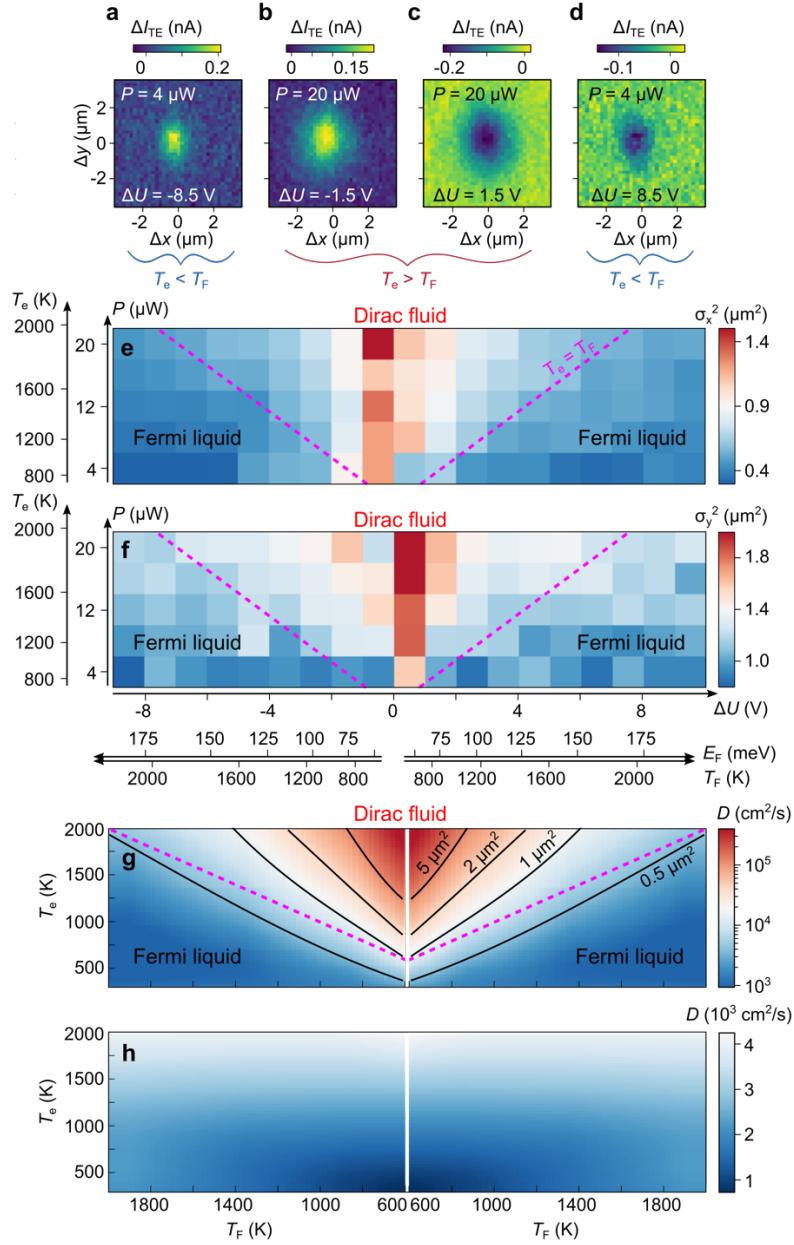

**Fig. 3. Fermi-liquid to Dirac-fluid crossover.** (**a**-**d**) Time-zero spatial maps of $\Delta I_{TE}$ for low optical power $P$ and high gate voltage $\Delta U$ (**a**,**d**), and vice versa (**b**,**c**), for *np*- and *pn*- junction (**a**-**b** and **c**-**d**, respectively). For larger ratio $T_e/T_F$ (*i.e.* larger $P/\Delta U$) the spatial extent is clearly larger. (**e**, **f**) Time-zero Gaussian widths for spatial scans with one pulse on the junction and the second one scanning across (**e**) and along (**f**) the graphene *pn*-junction, as a function of $P$ and $\Delta U$. The red dashed line shows the theoretical crossover temperature from Fermi liquid to Dirac fluid regime according to Ref. [14], thus showing our ability to controllably transition into the Dirac-fluid regime with strongly increase thermal diffusivity. (**g**, **h**) Calculation of the thermal diffusivity following Refs. [9,18] with only electron-electron interactions (**g**) and only long-range Coulomb scattering (**h**). The contours in (**g**) are the calculated time-zero spreads $\sigma^2_{calc}$ (see Methods).

14